\documentclass[sigconf, 11pt ]{acmart}

\usepackage{booktabs} 
\PassOptionsToPackage{hyphens}{url}\usepackage{hyperref}
\usepackage[anythingbreaks]{breakurl}

\setcopyright{none}

\acmConference[]{Presented as a poster at the 2017 Workshop on Fairness, Accountability, and Transparency in Machine Learning (FAT/ML 2017)}{}{} 
\acmYear{2017}

\begin{document}
\title{Is it ethical to avoid error analysis?}

\author{Eva Garc\'ia-Mart\'in}
\orcid{0000-0003-4973-9255}
\affiliation{%
  \institution{Blekinge Institute of Technology}
  \department{Dept. of Computer Science and Engineering}
  \city{Karlskrona} 
  \country{Sweden} 
  \postcode{371 79}
}
\email{eva.garcia.martin@bth.se}

\author{Niklas Lavesson}
\affiliation{%
  \institution{Blekinge Institute of Technology}
   \department{Dept. of Computer Science and Engineering}
  \city{Karlskrona} 
  \state{Sweden} 
  \postcode{371 79}
}
\email{niklas.lavesson@bth.se}

\renewcommand{\shortauthors}{E. Garc\'ia-Mart\'in et al.}

\begin{abstract}

Machine learning algorithms tend to create more accurate models with the availability of large datasets. In some cases, highly accurate models can hide the presence of bias in the data.
There are several studies published that tackle the development of discriminatory-aware machine learning algorithms. We center on the further evaluation of machine learning models by doing error analysis, to understand under what conditions the model is not working as expected. 
We focus on the ethical implications of avoiding error analysis, from a falsification of results and discrimination perspective.  Finally, we show different ways to approach error analysis in non-interpretable machine learning algorithms such as deep learning.

\end{abstract}

%
%


\keywords{Fairness, Transparency, Ethical Machine Learning, Error analysis}

\maketitle

\section{Introduction}

Fairness in machine learning studies the discriminatory impact of different machine learning algorithms, techniques or approaches from three different angles:  fairness, transparency and accountability. Fairness studies discrimination-aware data mining~\cite{zliobaite2015survey}, and centers on developing systems that either analyze if models are producing biased predictions, or develops systems that are discrimination-conscious-by-design~\cite{hajian2016algorithmic}.
As a clarification, when mentioning discrimination we refer to: "the unjust or prejudicial treatment of different categories of people"\footnote{\url{https://en.oxforddictionaries.com/definition/discrimination}}.

Big data and deep learning present ethical issues~\cite{hardt2014big}. Deep learning algorithms develop models that resemble a black box, being hard or impossible to interpret. Thus, making it very complicated if not impossible to understand the reasons behind the model's predictions. 
Big data may pose ethical issues, in particular in relation to the analysis of social data at a granular level. 
The model can produce biased decisions, but since it is built on large-scale datasets, that bias can remain unnoticed due to high predictive performance results~\cite{wallach2014big}.

This particular case is studied in this paper. We study what is called error analysis or model analysis. Error analysis addresses the reasons for the model to output an error, and if that error is due to standard noise or due to algorithmic bias or discrimination. We consider standard noise to be prediction errors in the testing phase, due to the model not being able to correctly classify a specific instance. Error analysis is directly connected to transparency and accountability, since it studies if a model is working as intended, giving also reasons why a model is outputting certain results. 
Since this is usually not possible in deep learning algorithms, we address if error or model analysis can be achieved in deep learning and what are the ethical issues behind it.

The goal of this study is to investigate the following question: Is it unethical to avoid error analysis of the model and the results?
There are several studies that address post-processing model approaches~\cite{hajian2016algorithmic} by focusing on discrimination-aware models~\cite{kamiran2010discrimination,hajian2015discrimination,kamiran2012decision}. One of these post-processing model approaches is to enforce further analysis of the model to understand how biased/unbiased it is. 
We believe that avoiding error analysis can present two ethical issues:

\begin{itemize}
  \item Falsification of results
  \item Discrimination
\end{itemize}

While the most commonly referred meaning of falsification in research is to disprove a proposition, hypothesis, or theory, in this study we focus on falsification as a form of research fraud.
In research ethics, falsification of results is also omitting relevant data and results intentionally. In this case, we argue that not investigating further the results of a model or the reasons behind its behavior is omitting part of the results. Since there has been an intent to omit part of the results, this has a direct link to research fraud, thus presenting an ethical concern. 
Regarding discrimination, feeding a model with unbalanced data can lead to the model outputting discriminative predictions. If the researcher is unaware of the type of data that the model is being trained on, a discriminative situation can occur. In this case, the model could be correct, since the reasons for these predictions is on the wrongly acquisition of biased and unbalanced data. One approach to avoid this is by error analysis. 


In the following sections we present the cases of falsification and discrimination connected to error analysis in machine learning. 
We study how error analysis is related to deep learning, if it can be achieved, and the ethical aspects behind it. Finally, we show the benefits from a research perspective of digging deep into a model's understanding.

\section{Falsification}

The form of falsification studied in this paper stands for: "Manipulating research materials, images, data, equipment, or processes. Falsification includes changing or omitting data or results in such a way that the research is not accurately represented. A person might falsify data to make it fit with the desired end result of a study\footnote{\url{http://ori.hhs.gov/definition-misconduct}}".

Is it really ethical to not disclose all the information regarding an algorithm, experiment, or model? From an ethical perspective, presenting a work that is not fully tested is not considered a good practice. However, in legal terms, it depends if there was intent from the researcher side to tamper with results, or hide evidence. 
If the intent exists, then it is considered falsification and research fraud. 
On the other hand, there could be cases where the researcher chooses not to work further on the model evaluation aspect, but with no intent to falsify results or data. While this could be regarded as unethical, it is not research fraud.  

One example of falsification is when researchers publish experiments only from datasets where the algorithm outputs highly accurate results, and omits those datasets where the algorithm fails. 
This is commonly known as cherry picking\footnote{\url{https://en.wikipedia.org/wiki/Cherry_picking}}, and it is directly falsification, since the researchers are intentionally omitting results to improve the chances of getting published by faking the performance of their algorithm or solution. 

In relation to avoiding error analysis, one can argue that it is also a case of falsification if intent can be proven, since the research and the results are not accurately presented. A good research ethics practice is to have a clear understanding of the results, ensuring an accurate portray of reality.
Confirmation bias plays an important part on this aspect. Confirmation bias is: "the tendency to search for, interpret, favor, and recall information in a way that confirms one's preexisting beliefs or hypotheses\footnote{\url{https://en.wikipedia.org/wiki/Confirmation_bias}}". 

A possible reason for a researcher to not proceed further into analyzing the results of the model could be because they obtained positive results, confirming their own hypothesis of the correctness of their work. 
This is also confirmation bias. As explained by Elsevier: "a person might falsify data to make it fit with the desired end result of a study\footnote{\url{https://www.publishingcampus.elsevier.com/websites/elsevier_publishingcampus/files/Guides/Quick_guide_RF02_ENG_2015.pdf}}".
In terms of research, not doing error analysis might lead to a model working differently than what is published, with the researcher being unaware of this. 
For example, by testing algorithm $A$ against algorithm $B$ in only one dataset, the results could be due to pure chance.

Our main argument is that the lack of understanding of the model and the results is an ethical concern. 
While only omitting results intentionally is considered as falsification and fraud, a lack of understanding of the model is unethical, unprofessional, and a bad research practice.

\section{Discrimination}
Big Data and discrimination are fully related nowadays. Big Data studies patterns in data at a granular level, thus reaching a level that can be key to the privacy of the person.
{\v{Z}}liobait{\.e}~\cite{zliobaite2015survey} gives a clear explanation of how non-discrimination can be defined in the context of machine learning: "(i) people that are similar in terms non-protected characteristics should receive similar predictions, and (ii) differences in predictions across groups of people can only be as large as justified by non-protected characteristics."
The first part of the definition means that the model should output the same value, in presence or not of the non-protected characteristic. 
A non-protected characteristic can be race or gender. The second part means that if (i) does not occur, the researcher should be able to explain the reasons behind it, they should be justified. 

In relation to model analysis, not doing model analysis can lead to a discriminatory and biased model.  
Wallach and Hardt portray a very clear example of when error analysis can be crucial~\cite{wallach2014big,hardt2014big}. Imagine a situation where there is a model that classifies user names as "real" or "fake" with a 95\% accuracy.
At a first glance, any researcher would be satisfied with the results. 
However, analyzing the results and the model carefully, they observed that the model outputs a high accuracy when it classifies names from American people, but it only achieves a 50\% accuracy when classifying names from other cultures. In this case, the model is correct, since it has generalized from the training data and built a model accordingly. The problem is the unbalanced and biased data that is used for training. If the model is trained with data that is mostly populated with American names, and almost no names from other cultures, then the model will output biased predictions. This can be avoided by further evaluating the model.

We argue that the researcher has the ethical obligation to ensure that their model is not biased and discriminatory if it is going to be applied on human-related scenarios. 
Several publications check for biases in different models. For example, Tramer et.al created a toolkit to identify biases in machine learning models~\cite{tramer2015discovering}. The consequences of not checking for this type of biases can lead to cases such as the ones reported where Google was labeling black people as gorillas~\cite{guynn2015google}. Or the case with the Microsoft chatbot "Tay"~\cite{taybot}.
Overall, researchers should build discrimination-aware models, ensuring their published models are working as expected by doing error and model analysis. 


\section{Deep Learning}

There are two key questions regarding deep learning and error analysis: 1) can we achieve error analysis in deep learning?, and 2) what are the ethical issues if we can not do error analysis in deep learning?

Deep learning algorithms are designed as black boxes, where very little information can be extracted from the models. This characteristic has been described as "opacity"~\cite{burrell2016machine}. In order to understand if conducting error analysis on these algorithms is possible or not, we need to first understand the reasons behind this opacity, and if there is a way out of it that can yield light into improving the transparency of the model. 

There are several forms of opacity. 
In the following paragraphs we see how they relate to error analysis, and if it can be achieved in such scenarios. 
The first form of opacity is due to \textit{intentional corporate or state secrecy}. 
Several companies have algorithms that are responsible for the company making big revenues, thus these companies do not want to make those algorithms public. One way to solve this issue is by promoting open source publishing of code, a practice that is quite common in top conferences and journals in data mining and machine learning. In relation to error analysis, \textit{intentional corporate or state secrecy} does not prevent the researchers of the algorithm to dig deeper into the models characteristics. So there is no conflict in this aspect.

The second form of opacity is due to \textit{a mismatch between mathematical procedures of machine learning algorithms and human styles of semantic interpretation}~\cite{burrell2016machine}. In other words, humans have trouble understanding them. This is directly an issue for error analysis and transparency. Since it is hard to interpret how the algorithm is building the model and making the decisions, we can not be transparent regarding the logic, results, and the reasons for those results. When trying to interpret the logic behind this type of models, we observe how different human-logic from machine-logic is, in particular in such high dimensional data~\cite{domingos2012few}. In light of this constraint, one suggestion to improve interpretability and transparency is to evaluate the discriminatory impact of the model, without actually knowing how the model works~\cite{dwork2012fairness}. 
For example, we could input certain type of biased data and observe the model's accuracy. Since we can not know the "gears" of the algorithm, we suggest to test the output given specific input.

This lack of interpretability arises many ethical concerns, in particular in key research fields like medicine and health. One can not explain the reasons behind the model, but the best approach that a researcher can follow is to increase the effort spent on evaluating a non-interpretable machine learning model to ensure that is not discriminatory and biased.

\section{Benefits of Error Analysis}

On a more general focus, error analysis has clear benefits apart from the ethical ones already mentioned. Publishing work that has been thoroughly evaluated, where the code is open source and available for everyone, and where there is a deep understanding of how it works, moves towards high quality research. 

On the other hand, cherry-picking, not digging deep into the model's behavior will produce lower quality research, that does not really help advance towards a more fair and transparent society. 

To finally motivate in favor of doing error analysis, there are four key points that are beneficial for every researcher and society in general: i) better understanding of the model, ii) transparency, iii) accountability, and iv) advancement of further research. 

\vspace{0pt}

\section{Take Away Notes}

This paper has addressed the issue of avoiding doing error analysis and its ethical concerns. We have argued that while only if researchers had the intent to omit results and do further analysis to favor their results is considered as falsification and fraud, omitting results and avoiding further model evaluation and analysis is unethical, unprofessional and not a good research practice.


Avoiding doing error analysis can also lead to a situation where the model is trained with biased data, creating a model that outputs discriminative predictions, such as what happened with Google labeling black people as Gorillas~\cite{guynn2015google}. There needs to be an understanding of the limitations of a model. An algorithm creates a model based on a mathematical function and training data. If that data is not accurately representing the reality, then the model produced will portray that non-accurate reality. Thus, the opened question is: \textit{how can we design AI that gives morally, justifiable, thoughtful, emphatic and/or fair responses?}

\begin{acks}
This work is part of the research project "Scalable resource-efficient systems for big data analytics" funded by the Knowledge Foundation (grant: 20140032) in Sweden.

\end{acks}

\bibliographystyle{ACM-Reference-Format}
\bibliography{lib} 

\end{document}